\documentclass[12pt]{article}

\usepackage{amssymb}
\usepackage{amsmath}
\usepackage{amscd}
\usepackage{latexsym}

\newcommand{\be}{\begin{equation}}
\newcommand{\ee}{\end{equation}}

\newcommand{\dlt}{\delta}

\newcommand{\bt}{\beta}
\newcommand{\rmM}{{\rm M}}

\newcommand{\al}{\alpha}
\newcommand{\ra}{\rightarrow}

\newcommand{\cM}{{\cal M}}
\newcommand{\cA}{{\cal A}}
\newcommand{\cL}{{\cal L}}

\begin{document}

\begin{center}

{\Large {\bf Quantum decision theory as quantum theory of
measurement} \\ [5mm]
V.I. Yukalov$^{1,2}$ and D. Sornette$^{1}$ } \\ [3mm]
{\it
$^1$Department of Management, Technology and Economics, \\
Swiss Federal Institute of Technology, Z\"urich CH-8032,
Switzerland \\ [3mm]

$^2$Bogolubov Laboratory of Theoretical Physics, \\
Joint Institute for Nuclear Research, Dubna 141980, Russia}

\end{center}

\vskip 1cm

\begin{abstract}

We present a general theory of  quantum information processing
devices, that can be applied to human decision makers, to atomic
multimode registers, or to molecular high-spin registers. Our
quantum decision theory is a generalization of the quantum theory
of measurement, endowed with an action ring, a prospect lattice and
a probability operator measure. The algebra of probability operators
plays the role of the algebra of local observables. Because of the
composite nature of prospects and of the entangling properties of
the probability operators, quantum interference terms appear, which
make actions noncommutative and the prospect probabilities non-additive.
The theory provides the basis for explaining a variety of paradoxes
typical of the application of classical utility theory to real human
decision making. The principal advantage of our approach is that it
is formulated as a self-consistent mathematical theory, which allows
us to explain not just one effect but actually all known paradoxes
in human decision making. Being general, the approach can serve as
a tool for characterizing quantum information processing by means
of atomic, molecular, and condensed-matter systems.

\end{abstract}

\vskip 2cm


\vskip 2cm


\newpage

\section{Introduction}

The classical theory of decision making is based on the expected utility
theory formalized by Von Neumann and Morgenstern [1]. In spite of its
normative appeal, researchers have uncovered several types of widespread
systematic violations of expected utility theory and of its underlying
assumptions, in experiments involving real human beings. Beginning with
the observations by Allais [2], Edwards [3], and Ellsberg [4] about fifty
years ago and continuing through the present, a growing body of clear
evidence has been accumulated, which shows that real-life decision makers
do not conform to many of the key assumptions and predictions of expected
utility theory. The deviations have been found to be systematic and
reproducible, and have been classified in so-called paradoxes and
fallacies, documented in a voluminous literature (see review articles
[5--7] and reference therein). While classical utility theory is perfectly
self-consistent, it is just that real human beings do not conform to its
predictions, leading to behaviors which appear paradoxical  {\it when
interpreted in the framework of classical utility theory}.

Here, we present an alternative ``Quantum decision theory'' (QDT) which,
while also being internally consistent as is classical utility theory,
provides correct predictions concerning the decisions made by real
humans under uncertainty. The behaviors of human beings, which appeared
paradoxical when interpreted with classical utility theory, find a simple
and unambiguous explanation, when interpreted through the lenses of QDT.
The underlying mechanisms at work to explain the characteristics of
decision making performed by real humans involve entanglement of actions,
the ordering of prospects with respect to the probabilities, the
interference between actions, and the mapping of the quantum interference
terms to human aversion towards loss and uncertainty.

The roots of our theory are found in quantum theory, based on the
mathematical techniques developed by Von Neumann [8] and Benioff [9,10].
Benioff [9,10] showed how the mathematical structure of quantum mechanics
provides a general description of quantum measurements [8] and of quantum
information processing [9,10]. Long before these works [8--10], Bohr himself
[11,12] mentioned that mental processes in many respects are similar
to those in quantum physics, and that quantum theory could serve
as a tool for attacking the problems associated with human thinking.

It is worth emphasizing that employing the mathematical techniques of
quantum theory for the description of human decision making does not
necessarily imply that psychological processes are truly quantum. This
simply means that such mathematical techniques are convenient tools for
formalizing the description of decision making, and provide novel testable
predictions.  The situation here could be compared with the following
well-known history of differential and integral calculi, which were
developed initially for describing the motion of planets under the
influence of gravity. But now, these calculi are used everywhere, for
problems having no relation to gravity. In the same way, the techniques
of quantum theory are nothing but the mathematical language of functional
analysis, which can be employed outside the quantum world, when
appropriate.

The aim of the present paper is to show that quantum decision theory
can be developed in such a way that it can be applicable not only to the
theory of measurements and to quantum information processing, but also to
some complex macroscopic problems, in particular by providing a coherent
structure to describe and predict decision making by real humans. For
this purpose, we reformulate the theory by accepting as its basic object,
not a manifold of simple actions but, a set of composite prospects. It
is the prospect compositeness that yields several nontrivial consequences,
such as decision interference, which reflects the presence of biases (with
respect to classical utility theory)  in the decision making process. The
developed QDT is shown to be mathematically self-consistent, and it can
explain the paradoxes of classical utility theory by accounting for the
empirical observations on decision making by real human beings. Its
usefulness is ultimately validated by its prediction of novel behaviors
that can be tested empirically. Our approach can also be applied to
quantum information processing on the manifold of composite prospects.
In order to show that QDT is mathematically justified, we present it
in a rigorous axiomatic way. Detailed calculations to explain the many
paradoxes reported in the framework of classical utility theory are given
in [7,16].

\section{Basic definitions}

The simplest elements in decision making theory are intended actions, or
just actions. We shall denote actions as $A_n$, enumerating them with the
index $n=1,2,\ldots$.

\vskip 2mm
{\bf Definition 1}.   {\it The action ring is a set $\cA=\{ A_n: \;
n=1,2,\ldots\}$ of actions $A_n$, endowed with two binary operations,
the associative reversible addition, denoted as $A_m+A_n$, and the
distributive noncommutative multiplication, denoted as $A_mA_n$, which
possesses a zero element $0$, called the empty action, such that
$A_n0=0A_n=0$.}

\vskip 2mm
{\bf Definition 2}. {\it Two actions $A_m$ and $A_n$ from $\cA$, with
$m\neq n$, are termed disjoint if and only if they are divisors of zero
with respect to each other, i.e., $A_mA_n=0$.}

\vskip 2mm
{\bf Definition 3}. {\it An action $A_n\in\cA$ is composite, if and only
if it can be represented as a union}
\be
\label{eq1}
A_n = \bigcup_{j=1}^{{\rm M}_n} A_{nj}
\ee
{\it of ${\rm M}_n>1$ disjoint subactions $A_{nj}\in\cA$, called action
modes, and $A_n$ is simple if it cannot be decomposed into a union of
form (\ref{eq1})}.

\vskip 2mm
{\bf Definition 4}.  {\it An action prospect $\pi_n$ is a conjunction}
\be
\label{eq2}
\pi_n = \bigcap_j A_{n_j} \qquad ( A_{n_j}\in\cA_n \subset \cA )
\ee
{\it of the actions $A_{n_j}$ taken from a subset $\cA_n=\{ A_{n_j}:\;
n_j=1,2,\ldots\}$ of the action ring $\cA$}.

\vskip 2mm
{\bf Definition 5}. {\it A prospect $\pi_n$ is composite if at least one
of the actions $A_{n_j}$ in conjunction (\ref{eq2}) is composite, and the
prospect $\pi_n$ is simple when all actions $A_{n_j}$ in conjunction
(\ref{eq2}) are simple}.

\vskip 2mm
{\bf Definition 6}. {\it Elementary prospects $e_\al$, with $\al=1,2,\ldots$,
are simple prospects}
\be
\label{eq3}
e_\al = \bigcap_j A_{\al_j} \qquad
(A_{\al_j} \in \cA_\al \subset \cA) \; ,
\ee
{\it such that any two elementary prospects $e_\al$ and $e_\bt$, with
$\al\neq\bt$, are mutually disjoint},
\be
\label{eq4}
e_\al e_\bt = 0 \qquad (\al\neq \bt) \; .
\ee

\vskip 2mm
{\bf Definition 7}. {\it The prospect lattice $\cL$ is a set $\{\pi_n\}$
of partially ordered prospects $\pi_n$, such that any pair of prospects
can be ordered according to one of the linear transitive binary ordering
relations $\{ <,=,\leq\}$. Therefore, for any two prospects $\pi_1$ and
$\pi_2$, one has either $\pi_1<\pi_2$, or $\pi_1>\pi_2$, or $\pi_1=\pi_2$,
or $\pi_1\leq\pi_2$, or $\pi_1\geq\pi_2$. The minimal element is the empty
action $0$, and there exists a maximal element $\pi_*$, which makes the
lattice
\be
\label{eq5}
\cL = \{ \pi_n: \; 0 \leq \pi_n \leq \pi_* \}
\ee
complete}.

\vskip 2mm
{\bf Remark}: The ordering operations are performed by using the
probabilities of the prospects introduced in Definition 18.

\vskip 2mm
{\bf Definition 8}. {\it For each mode $A_{nj}$ of an action $A_n$ there
corresponds a function $\cA\ra\mathbb{C}$, called the mode state $|A_{nj}>$,
such that it possesses a Hermitian conjugate state $<A_{nj}|$ and, for
any two mode states $|A_{ni}>$ and $|A_{nj}>$, a scalar product
\be
\label{eq6}
< A_{ni} | A_{nj} > \; = \; \dlt_{ij}
\ee
is defined}.

\vskip 2mm
{\bf Definition 9}.  {\it The mode space is the Hilbert space}
\be
\label{eq7}
\cM_n \equiv {\rm Span} \; \{ | A_{nj} > : \;
j=1,2,\ldots,\rmM_n \} \; ,
\ee
{\it which is the closed linear envelope spanning all mode states}.

\vskip 2mm
{\bf Definition 10}. {\it For each elementary prospect $e_\al$, there
corresponds a function $\cA\times\cA\times\ldots\times\cA\ra\mathbb{C}$,
called the basic state}
\be
\label{eq8}
| e_\al>\; \equiv \; | A_{\al_1} A_{\al_2} \ldots > \;
\equiv \; \bigotimes_j | A_{\al_j} > \; ,
\ee
{\it such that it has a Hermitian conjugate state $< e_\al|$ and, for any
two basic states $|e_\al>$ and $|e_\bt>$, there exists the scalar product}
\be
\label{eq9}
< e_\al | e_\bt > \; \equiv \; \prod_j < A_{\al_j} A_{\bt_j} > \; =
\; \prod_j \dlt_{\al_j \bt_j} \equiv \dlt_{\al\bt} \; .
\ee

\vskip 2mm
{\bf Definition 11}.  {\it The mind space is the Hilbert space}
\be
\label{eq10}
\cM \equiv {\rm Span} \; \{ | e_\al> \} = \bigotimes_n \cM_n \; ,
\ee
{\it which is the closed linear envelope spanning all basic states}.

\vskip 2mm
{\bf Definition 12}. {\it For each prospect $\pi_n$, there corresponds
a vector $|\pi_n>$ in the mind space $\cM$, named the prospect
state $|\pi_n>\in\cM$, possessing a Hermitian conjugate $<\pi_n|$}.

\vskip 2mm
{\bf Definition 13}.  {\it  The vacuum state is the vector $|0>\in\cM$
corresponding to the empty prospect, for which}
\be
\label{eq11}
< \pi_n | 0 > \; = \; < 0 |\pi_n> \; = \; 0 \; ,
\ee
{\it for any $|\pi_n>\in\cM$}.

\vskip 2mm
{\bf Definition 14}.  {\it A state of mind is a given specific vector
$|\psi>\in\cM$ from the mind space, which is normalized, such that}
\be
\label{eq12}
< \psi | \psi > \; = \; 1 \; .
\ee

\vskip 2mm
{\bf Definition 15}. {\it The probability operator for a prospect
$\pi_n\in\cL$ is the self-adjoint operator}
\be
\label{eq13}
\hat P(\pi_n) \equiv | \pi_n > < \pi_n | \; ,
\ee
{\it defined on $\cM$ and satisfying the normalization condition}
\be
\label{eq14}
\sum_n < \psi | \hat P(\pi_n) | \psi > \;  =  1 \; ,
\ee
{\it in which the summation is over all $\pi_n\in\cL$}. 

\vskip 2mm
{\bf Definition 16}. {\it The algebra of probability operators is the
involutive bijective algebra}
\be
\label{eq15}
{\cal P} \equiv \{ \hat P(\pi_n): \; \pi_n\in \cL \} \; ,
\ee
{\it with the bijective involution given by the Hermitian conjugation
${\cal P}^+={\cal P}$}.

\vskip 2mm
{\bf Definition 17}. {\it The expectation value of a probability operator
$\hat P(\pi_n)$, under the state of mind $|\psi>\in\cM$, is the average
defined by}
\be
\label{eq16}
< \hat P(\pi_n) > \; \equiv \; < \psi | \hat P(\pi_n) | \psi > \; .
\ee

\vskip 2mm
{\bf Definition 18}. {\it The prospect probability $p(\pi_n)$ of a prospect
$\pi_n\in\cL$, under the state of mind $|\psi>\in\cM$, is the expectation
value}
\be
\label{eq17}
p(\pi_n) \; \equiv < \hat P(\pi_n) > \; ,
\ee
{\it with the normalization condition}
\be
\label{eq18}
\sum_n p(\pi_n) = 1 \; ,
\ee
{\it where the summation is over all $\pi_n\in\cL$}.

\vskip 2mm
{\bf Definition 19}. {\it The probabilistic state is the set}
\be
\label{eq19}
< {\cal P} > \; = \; \{ p(\pi_n) : \; \pi_n\in \cL \}
\ee
{\it of the prospect probabilities $p(\pi_n)$ for all $\pi_n\in\cL$}.

\vskip 2mm
{\bf Definition 20}. {\it Two prospects, $\pi_1$ and $\pi_2$ from $\cL$, are
indifferent if and only if}
\be
\label{eq20}
p(\pi_1) = p(\pi_2) \qquad (\pi_1=\pi_2) \; .
\ee

\vskip 2mm
{\bf Definition 21}. {\it Between two prospects, $\pi_1$ and $\pi_2$ from
$\cL$, the prospect $\pi_1$ is preferred to $\pi_2$ if and only if}
\be
\label{eq21}
p(\pi_1) > p(\pi_2) \qquad (\pi_1 > \pi_2) \; .
\ee

\vskip 2mm
{\bf Definition 22}. {\it The prospect $\pi_*$, which is the maximal
element of $\cL$, is optimal if and only if the related prospect probability
is the largest among all $p(\pi_n)$, so that}
\be
\label{eq22}
p(\pi_*) = \sup_n p(\pi_n) \qquad (\pi_* =\sup_n \pi_n ) \; .
\ee

\vskip 2mm
{\bf Definition 23}. {\it The binary conjunction set}
\be
\label{eq23}
{\cal B} \equiv \{ \pi_n e_\al : \;
\pi_n \in \cL \; , e_\al\in \cA \}
\ee
{\it is the family of the conjunction actions $\pi_n e_\al\in\cA$}.

\vskip 2mm
{\bf Definition 24}. {\it The probability operator for the conjunction action
$\pi_n e_\al\in {\cal B}$ is the self-adjoint operator}
\be
\label{eq24}
\hat P(\pi_n e_\al) \equiv \hat P(e_\al) \hat P(\pi_n)
\hat P(e_\al) \; ,
\ee
{\it defined on $\cM$ and satisfying the normalization condition}
\be
\label{eq25}
\sum_{n, \al} \hat P(\pi_n e_\al) = \hat 1 \; ,
\ee
{\it in which $\hat 1$ is the identity operator on $\cM$ and the summation
is over all $\pi_n\in\cL$ and $e_\al\in\cA$}.

\vskip 2mm
{\bf Definition 25}. {\it The probability $p(\pi_n e_\al)$ of the conjunction
action $\pi_n e_\al\in{\cal B}$, under the state of mind $|\psi>\in\cM$,
is the expectation value}
\be
\label{eq26}
p(\pi_n e_\al) \; \equiv \; < \hat P(\pi_n e_\al) > \; ,
\ee
{\it with the normalization condition}
\be
\label{eq27}
\sum_{n, \al} p(\pi_n e_\al) = 1 \; ,
\ee
{\it where the summation is over all $\pi_n e_\al\in{\cal B}$.}

{\bf Remark}:
The probability $p(\pi_n e_\al)$ is the analog of the classical utility,
in the sense that, without additional effects, the conjunction action
which would be taken corresponds to the largest $p(\pi_n e_\al)$.
The next definition shows that there are additional interference terms,
which modify this classical picture.

\vskip 2mm
{\bf Definition 26}. {\it The quantum interference term for a prospect
$\pi_n\in\cL$, under the state of mind $|\psi>\in\cM$, is the average}
\be
\label{eq28}
q(\pi_n) \equiv \sum_{\al\neq\bt} < \hat P(e_\al) \hat P(\pi_n)
\hat P(e_\bt) > \; ,
\ee
{\it in which the summation is over all $e_\al\in\cA$ and $e_\bt\in\cA$}.

\vskip 2mm
{\bf Definition 27}. {\it Two prospects, $\pi_1$ and $\pi_2$ from $\cL$,
are equally repulsive (equally attractive) if and only if}
\be
\label{eq29}
q(\pi_1) =  q(\pi_2) \; .
\ee

\vskip 2mm
{\bf Definition 28}. {\it Between two prospects, $\pi_1$ and $\pi_2$ from
$\cL$, the prospect $\pi_1$ is more repulsive (less attractive) if and only
if}
\be
\label{eq30}
q(\pi_1) < q(\pi_2) \; .
\ee

\vskip 2mm
{\bf Definition 29}. {\it A prospect $\pi_1\in\cL$ is more repulsive (less
attractive) than $\pi_2\in\cL$, in the sense of inequality (\ref{eq30}),
when one of the following is true:
\begin{itemize}
\item
$\pi_1$ leads to more uncertain gains than $\pi_2$ (less certain gain),

\item
$\pi_1$ gives a more certain loss than $\pi_2$  (less uncertain loss),
\end{itemize}
or $\pi_1$, as compared to $\pi_2$, requires to be:
\begin{itemize}
\item
more active under uncertainty (less passive under uncertainty),

\item
more passive under certainty (less active under certainty).
\end{itemize} }

{\bf Remark}: This classification is based on empirical evidence on human
decision making. To give a more complete and precize description, we formulate
in the definition below the counterpart of Definition 29. 

\vskip 2mm
{\bf Definition 30}.  {\it A prospect $\pi_2\in\cL$ is more attractive (less
repulsive) than $\pi_1\in\cL$, in the sense of inequality (\ref{eq30}), when
one of the following is true:
\begin{itemize}
\item
$\pi_2$ leads to less uncertain gains than $\pi_1$ (more certain gain),

\item
$\pi_2$ gives a less certain loss than $\pi_1$  (more uncertain loss),
\end{itemize}
or $\pi_2$, as compared to $\pi_1$, requires to be:
\begin{itemize}
\item
less active under uncertainty (more passive under uncertainty),

\item
less passive under certainty (more active under certainty).
\end{itemize}   }

\section{Main results}

The above definitions make it possible to develop a self-consistent
quantum decision theory (QDT). This theory generalizes the quantum
theory of measurement developed by Von Neumann [8] and Benioff [9,10].
The principal innovation of our QDT is that it deals with composite
prospects, and not with simple actions. Also, the probability operator
(\ref{eq13}), generally, is an entangling operator [13--15]. These
features of our QDT result in the appearance of the quantum interference
term (\ref{eq28}) and, as a consequence, in the occurrence of interferences
between decisions. The latter serve as measures characterizing the prospects
as more or less repulsive (more or less attractive). The novel quantity,
the quantum interference term (\ref{eq28}), makes the QDT fundamentally
different from classical utility theory. In the frame of QDT, all known
paradoxes, plaguing classical utility theory, disappear. In other words,
QDT provides a coherent and realistic description of real human decision
making processes. Below, we present the main theorems of QDT, which make
clear that the theory is self-consistent and explain how the paradoxes
disappear within QDT. The detailed discussion and proofs of these theorems
will be given elsewhere [16].

\vskip 2mm
{\bf Proposition 1}. {\it The probability $p(\pi_n)$ of a prospect
$\pi_n\in\cL$ is the sum}
\be
\label{eq31}
p(\pi_n) = \sum_\al p(\pi_n e_\al) \; + \; q(\pi_n)
\ee
{\it of the conjunction probabilities (\ref{eq26}) and the quantum
interference term (\ref{eq28})}.

\vskip 2mm
{\bf Proposition 2}. {\it The sum of the quantum interference terms
$q(\pi_n)$ over all $\pi_n\in\cL$ is zero},
\be
\label{eq32}
\sum_n q(\pi_n) = 0 \; .
\ee

\vskip 2mm
{\bf Proposition 3}. {\it Between two prospects, $\pi_1$ and $\pi_2$ from
$\cL$, the prospect $\pi_1$ is preferred to $\pi_2$ if and only if}
\be
\label{eq33}
\sum_\al \; [ p(\pi_1 e_\al) - p(\pi_2 e_\al) ] \; > \;
q(\pi_2) - q(\pi_1) \; .
\ee

This proposition 3 shows that the preference of prospect $\pi_1$ over
prospect $\pi_2$ depends not merely on the relations between the
probabilities $p(\pi_1e_\al)$ and $p(\pi_2e_\al)$ but also on the relation
between the quantum interference terms $q(\pi_1)$ and $q(\pi_2)$. These
terms $q(\pi_1)$ and $q(\pi_2)$  influence the decision, by accounting for
such emotional feelings as the prospect attraction and prospect repulsion
which are involved in the process of decision making. The preferred prospect
$\pi_n$ is the one whose combined utility ($\sum_\al p(\pi_n e_\al)$) and
attraction ($q(\pi_n)$) are such that they satisfy the inequality
(\ref{eq33}).

The criterion (\ref{eq33}) is the basis for explaining a variety
of paradoxes typical of the application of classical utility theory
to real human decision making  [7]. In classical utility theory, the
interference terms $q(\pi_n)$ are absent, so that the preferred prospect
always corresponds to the largest utility. Real human beings do not follow
this specification, as exemplified by  the Allais paradox [2], the Ellsberg
paradox [4], the Rabin paradox [17], the Kahneman-Tversky paradox [18],
disjunction effect and the conjunction fallacy [5--7], which appear for
decisions made under uncertainty. The novel quantum terms $q(\pi_n)$ in
QDT makes it possible to resolve all these paradoxes.

The possibility of relating the disjunction effect with the interference
of actions was suggested earlier (see, e.g., [19,20]). However, the
principal advantage of our approach is that it is formulated  as a
self-consistent mathematical theory, which allows us to explain not
just one effect but actually all known paradoxes.

Our quantum decision theory is nothing but a general quantum theory of
measurement, endowed with an action ring, a prospect lattice  and a
probability operator measure. The algebra of probability operators plays
the role of the algebra of local observables. Because of the composite
nature of prospects and of the entangling properties of the probability
operators, quantum interference terms appear, which make actions
noncommutative and the prospect probabilities nonadditive. The developed
theory is self-consistent, containing no paradoxes, in contrast to those
that plague classical utility theory.

Our approach can be applied both to the description of human decision
making [7] as well as to quantum information processing. The tools for
realizing the latter can be of different physical nature. For example,
a convenient tool for quantum information processing can be constructed
on the basis of an optical lattice with multimode nonground-state
condensates of neutral atoms in each lattice site [21,22]. Then each
mode state $|A_{nj}>$ corresponds to a topological coherent mode of type
$j$, generated in a lattice site $n$. The mode space (\ref{eq7}) is a
Hilbert space of coherent modes of an $n$-site. Basic states (\ref{eq8})
form a basis for the states of the overall optical lattice. The mind space
(\ref{eq10}) is the Hilbert space for the coherent states of the whole
optical lattice. The prospect state $|\pi_n>$ is a state of $\cM$, composed
of the superposition of the coherent modes generated in different sites.
The state of mind $|\psi>$ is a given reference state of $\cM$.
Consequently, the prospect probability (\ref{eq17}) is the squared modulus
$$
p(\pi_n) = | < \pi_n |\psi> |^2
$$
of the transition amplitude $<\pi_n|\psi>$. The topological coherent modes
can be generated by applying an external resonant alternating field, either
modulating the trapping potential or varying atomic interactions [23,24].
The generation of the coherent modes can be well regulated, providing the
possibility of creating an atomic multimode register [21,22].

Another way of constructing a quantum register is by using molecular
magnets composed of nanomolecules with high spins. The spin states of
each molecule correspond to the mode states, forming the mode space
(\ref{eq7}). The states of the whole molecular magnet pertain to the mind
space (\ref{eq10}). The spin states can be governed by applying external
magnetic fields. An ultrafast spin regulation can be achieved by coupling
the molecular magnet to a resonant electric circuit [25].

We have here presented the general theory of  quantum information
processing devices, that applies to human decision makers, to atomic
multimode registers, or to molecular high-spin registers, or to other
condensed-matter registers. The common procedure for any quantum
register, playing the role of a decision maker, starts with the
classification of action prospects and ordering them through the
evaluation of their probabilities. Finding the optimal prospect
serves as a key for the subsequent functioning of the
information-processing device. Technical and experimental details of
such a quantum information-processing operation, requires separate
investigations, depending on the physical nature of a particular
system, chosen as a quantum register. But in any case, the operation
of these devices can be based on the mathematical scheme presented
in this Letter.

\vskip 5mm

{\bf Acknowledgement}: One of the authors (V.I.Y.) is grateful to
P.A. Benioff for useful correspondence and helpful remarks. Fruitful
discussions with E.P. Yukalova are appreciated.


\vskip 1cm


\begin{thebibliography}{99}

\bibitem{1}
J. Von Neumann, O. Morgenstern, Theory of Games and Economic
Behavior, Princeton University, Princeton, 1953.

\bibitem{2}
M. Allais,  Econometrica 21 (1953) 503.

\bibitem{3}
W. Edwards, J. Exp. Psychol. 50 (1955) 201.

\bibitem{4}
D. Ellsberg,  Quart. J. Econom. 75 (1961) 643.

\bibitem{5}
A. Tversky, D. Kahneman,  Psychol. Rev. 90 (1983) 293.

\bibitem{6}
M.J. Machina,  in: New Palgrave Dictionary of Economics, Macmillan,
New York, 2008.

\bibitem{7}
V.I. Yukalov, D. Sornette, arXiv:0802.3597 (2008).

\bibitem{8}
J. Von Neumann, Mathematical Foundations of Quantum Mechanics,
Princeton University, Princeton, 1955.

\bibitem{9}
P.A. Benioff, J. Math. Phys. 13 (1972) 908.

\bibitem{10}
P.A. Benioff, J. Math. Phys. 13 (1972) 1347.

\bibitem{11}
N. Bohr, Naturwiss. 17 (1929) 483.

\bibitem{12}
N. Bohr, Erkenntniss. 6 (1937) 293.

\bibitem{13}
V.I. Yukalov, Phys. Rev. Lett. 90 (2003) 167905.

\bibitem{14}
V.I. Yukalov, Mod. Phys. Lett. B 17 (2003) 95.

\bibitem{15}
V.I. Yukalov, Phys. Rev. A 68 (2003) 022109.

\bibitem{16}
V.I. Yukalov, D. Sornette, arXiv:0808.0112 (2008).

\bibitem{17}
M. Rabin, Econometrica 68 (2000) 1281.

\bibitem{18}
D. Kahneman, A. Tversky, Econometrica 47 (1979) 263.

\bibitem{19}
J.R. Busemeyer, Z. Wang, J.T. Townsend, J. Math. Psychol. 50 (2006)
220.

\bibitem{20}
J.R. Busemeyer, M. Matthew, Z. Wang, in: Proc. Cogn. Sci. Soc.,
Washington, 2007.

\bibitem{21}
V.I. Yukalov, E.P. Yukalova, Laser Phys. 16 (2006) 354.

\bibitem{22}
V.I. Yukalov, E.P. Yukalova, Phys. Rev. A 73 (2006) 022335.

\bibitem{23}
V.I. Yukalov, E.P. Yukalova, V.S. Bagnato, Phys. Rev. A 56 (1997) 4845.

\bibitem{24}
V.I. Yukalov, E.P. Yukalova, V.S. Bagnato, Phys. Rev. A 66 (2002) 043602.

\bibitem{25}
V.I. Yukalov, V.K. Henner, P.V. Kharebov, Phys. Rev. B 77 (2008) 134427.

\end{thebibliography}
\end{document}